\documentclass[11pt,a4paper]{article}
\pdfoutput=1
\usepackage{jheppub}
\usepackage{graphicx}
\usepackage{lipsum}
\usepackage{color}
\usepackage{url}
\usepackage{amsmath,amssymb}
\usepackage{slashed}
\usepackage{mathtools}
\usepackage{xcolor}
\usepackage[all]{hypcap}
\allowdisplaybreaks

\def\beq#1\eeq{\begin{align}#1\end{align}}

\newcommand{\ZP}{Z^\prime}

\newcommand{\be}{\begin{equation}}
\newcommand{\ee}{\end{equation}}
\newcommand{\bea}{\begin{eqnarray}}
\newcommand{\eea}{\end{eqnarray}}
\newcommand{\smelli}{{\tt smelli2.3.2}}
\newcommand{\flavio}{{\tt flavio2.3.3}}

\begin{document}

\title{{\boldmath$M_W$} helps select {\boldmath$\ZP$} models for {\boldmath
    $b\rightarrow s \ell \ell$} anomalies }

\author[a]{Ben Allanach}
\emailAdd{B.C.Allanach@damtp.cam.ac.uk}
\author[b]{and Joe Davighi}
\emailAdd{joe.davighi@physik.uzh.ch}

\affiliation[a]{DAMTP, University of Cambridge, Wilberforce Road, Cambridge,
  CB3 0WA, United Kingdom}
\affiliation[b]{Physik-Institut, Universit\"at Z\"urich, CH--8057 Z\"urich, Switzerland}

\abstract{
  As shown in Ref.~\cite{Allanach:2021kzj}, the Third Family Hypercharge
  ($Y_3$) Model changes the Standard Model prediction for $M_W$ whilst
  simultaneously explaining anomalies in $b\to s\ell\ell$ transitions via a
  heavy $Z^\prime$ gauge boson which is spawned by a spontaneously broken
  gauged $U(1)_{Y_3}$ symmetry. The 2022 CDF II measurement of $M_W$, which is
  far from the Standard Model prediction in the statistical sense, somewhat disfavours the $Y_3$ model. Here, we generalise the gauge charge assignments to the anomaly-free combination $s Y_3 + t (B_3-L_3)$ and show that incorporating the 2022 CDF II measurement of $M_W$ selects a viable domain of integers $s$ and $t$. For example, $s=1, t=-3$ yields a $p-$value of .12 in a two-parameter global fit to 277 electroweak and flavour changing $b$ data, much improving a SM $p-$value of $5\times 10^{-6}$.
}

\maketitle

\section{Introduction \label{sec:int}}

Recently, the CDF II Collaboration reported a measurement of the $W$ boson
mass 
$M_W=80.4335\pm0.0094$ GeV~\cite{CDF:2022hxs}
that disagrees by many sigma with the Standard Model
(SM) prediction of the quantity. One average\footnote{For an electroweak fit
  in terms of $S$, $T$ and $U$ parameters, see Ref.~\cite{Lu:2022bgw}.} of all current measurements
yields~\cite{deBlas:2022hdk}
\begin{equation}
  M_W = 80.4133 \pm 0.0080 \text{~GeV}, \label{avMW}
\end{equation}
  still in significant disagreement (with a pull of -6.4$\sigma$)
with SM predictions; we call this disagreement the $M_W$
anomaly. Note that the top quark mass $m_t$ plays an important role in the
prediction of $M_W$ in the SM: throughout this paper, we impose the following constraint
upon the top quark pole mass,
\begin{equation}
m_t=171.79 \pm 0.38 \text{~GeV}, \label{avmt}
\end{equation}
based on a combination of the latest
  data~\cite{deBlas:2022hdk}. We use the
  \smelli\ defaults~\cite{Aebischer:2018iyb}
  for other electroweak constraints and parameters.  

There are also discrepancies between SM predictions and some measurements
of flavour-changing $B$-meson decays, which we collectively call the
$b\rightarrow s\ell\ell$ anomalies. For example, various lepton flavour
universality (LFU) observables like the ratios of branching ratios
$R_{K^{(\ast)}}=BR(B \rightarrow K^{(\ast)} \mu^+ \mu^-)/BR(B \rightarrow
K^{(\ast)} e^+ e^-)$ are observed to be lower than their SM predictions in
several channels and several different bins of di-lepton invariant
mass squared~\cite{LHCb:2017avl,LHCb:2019hip,LHCb:2021trn}. Similar double
ratio 
measurements in $B^0 \rightarrow K^0_s \ell^+ \ell^-$ and $B^+ \rightarrow
K^{\ast+} \ell^+ \ell^-$ decays~\cite{LHCb:2021lvy} are consistent with a
similar deficit in di-muons over di-electrons, albeit with lower statistics
than in $R_K^{(\ast)}$.
The
predictions for all aforementioned double ratios have rather small theoretical uncertainties
due to cancellations and the SM predictions are generically considered to be
robust in the di-lepton invariant mass squared bins of interest. $BR(B_s\rightarrow \mu^+\mu^-)$ also has quite small theoretical
uncertainties in its SM prediction. The combined measurements 
of $BR(B_s\rightarrow\mu^+\mu^-)$ are in a 2$\sigma$ tension with its SM prediction~\cite{ATLAS:2018cur,CMS:2013dcn,CMS:2014xfa,LHCb:2017rmj,LHCb:2021vsc}.
Several other $B-$meson decay observables appear to be in tension with SM predictions
even when their larger theoretical uncertainties are taken into account: for
example angular distributions in $B\rightarrow K^\ast \mu^+ \mu^-$
decays~\cite{LHCb:2013ghj,LHCb:2015svh,ATLAS:2018gqc,CMS:2017rzx,CMS:2015bcy,Bobeth:2017vxj},
and $BR(B_s \rightarrow \phi \mu^+
\mu^-)$~\cite{LHCb:2015wdu,CDF:2012qwd}. For these quantities though, there
is room for argument about the assumed size of the theoretical uncertainties
and indeed the best way of estimating the SM predictions for them.

\begin{table}\begin{center}
  \begin{tabular}{|c|ccc|c|c|}\hline
category & $n_{obs}$ & $\chi^2$ & $p$ & $p$(global)& pull($M_W$) \\ \hline
quarks & 224 & 262 & .043 & & \\
LFU FCNCs & 23 & 39.4 & .018 & & \\ \hline
EWPOs({\tt smelli}) & 31 & 38.4 & .17 & .0074 & -2.1\\ 
EWPOs(CDF II) & 30 & 93.2 & 2.2$\times 10^{-8}$ & $4.6\times 10^{-6}$ & -6.4 \\     
    \hline
  \end{tabular}
  \caption{\label{tab:SMpvals} SM goodness of fit as calculated by \smelli{}.
    $n_{obs}$ shows the number of observables in each category. $\chi^2$ denotes 
    the $\chi^2$ statistic within that category, $p$ is the $p-$value of the
    category, and $p$(global) is the global $p-$value of all observables.
    The category  `LFU FCNCs' contains lepton flavour universality violating
    flavour changing observables such as $R_K^{(\ast)}$ as well as $BR(B_s
    \rightarrow \mu^+ \mu^-)$, where theoretical uncertainties are relatively small.
    The `quarks' category contains other flavour-changing $b$ observables, some of which
    have large theoretical uncertainties. The category
    EWPOs({\tt  smelli}) contains
    electroweak precision observables for the default \smelli\ combination of
    $M_W$ (\ref{smelliMW}) which excludes the CDF II measurement, whereas
    EWPOs(CDF II) 
    includes it in a global average \'a la (\ref{avMW}),
    (\ref{avmt}). $p$(global) includes the `quarks' category, the `LFU FCNCs' 
    category and the `EWPO' category relevant for the respective
    combination of $M_W$ measurements. For a definition of the
    pull, see (\ref{pull}) and the discussion of it in the surrounding text.
  }
 \end{center}
  \end{table}
We display the statistical tensions in the SM due to the $M_W$ and
$b\rightarrow s \ell \ell$ anomalies in Table~\ref{tab:SMpvals}, as calculated
by the computer program\footnote{We use the development version of
  \smelli\ and its sub-program \flavio\ that were on {\tt github} 
  on 
27/4/22. The correlations between the various theoretical uncertainties were
re-calculated.} \smelli~\cite{Aebischer:2018iyb}. We see from the table that
the 
$b\rightarrow s \ell \ell$ anomalies disfavour the SM:
the $b\rightarrow s \ell \ell$ data yields a poor fit with a 
global $p-$value of .0074, even when 
using the default \smelli\ constraints upon the EWPOs (\ref{smelliMW}),
i.e.\ \emph{excluding} the new CDF II 
measurement of $M_W$ (although the EWPOs have an acceptable fit in and of
themselves with a
$p-$value of~.17). Taking into account the CDF II measurement of $M_W$ as in (\ref{avMW})
exacerbates an already poor quality of fit, lowering the global $p-$value to
$5 \time 10^{-6}$.  

Global fits find that new physics
contributions to the
$(\overline{b} \gamma^\mu P_L s) (\overline{\mu} \gamma_\mu P_X \mu)+H.c.$ 
vertex in the Lagrangian density can ameliorate the fit to the
$b\rightarrow s \ell \ell$
data~\cite{Alguero:2019ptt,Alok:2019ufo,Ciuchini:2019usw,Aebischer:2019mlg,Datta:2019zca,Kowalska:2019ley,Arbey:2019duh},
where $P_L$ is the left-handed projection operator. $P_X$ corresponds to the
helicity projection of the muon pair in the effective vertex: the fits agree
that a purely right-handed projection $P_X=P_R$ is
disfavoured, whereas a mixture in the domain $P_X\approx P_L$ to $P_X\approx P_L+P_R$
is preferred~\cite{Egede:2022rxc}. Whereas the fitting groups yield very
similar 
results when fitting just 
to the `LFU' category of observables, there are some differences observed when
the `quarks' category is included: for example, which option out of 
$P_X=P_L$ or $P_X=P_L+P_R$ has a better fit. Such differences in the
constraints can arise from the  
treatment of theoretical uncertainties in the `quarks' category. 
The lesson one learns from such studies is that, from a new physics point of
view in order to fit the $b\rightarrow s \ell \ell$ anomalies, one needs a new
physics state that couples to left-handed quark fields and left-handed muon fields
in a family non-universal manner (but it may or may not also couple to
right-handed muon fields and/or electron fields). According to Ref.~\cite{Altmannshofer:2021qrr}, there is a 
mild preference for a family universal coupling of new physics
to leptons
as well as a specific and different new physics contribution to the coupling
to di-muon pairs.

One category of new physics state that can have such couplings is a heavy
electrically-neutral vector boson, dubbed a
$Z^\prime$~\cite{Gauld:2013qba,Buras:2013dea,Buras:2013qja,Buras:2014yna,Allanach:2015gkd}. In
specific models, one obtains the $Z^\prime$ from a spontaneously broken
additional $U(1)_X$ gauge symmetry under which the SM fermions have family
dependent charges. In a consistent ultra-violet (UV) complete model, quantum field
theoretic anomalies should cancel\footnote{The equations for cancellation of local gauge anomalies have been
solved analytically for a gauge group of the form SM$\times U(1)_X$~\cite{Allanach:2020zna}, although
it is often easiest to search lists to identify
anomaly-free charge assignments of phenomenological interest~\cite{Allanach:2018vjg}. It has also been shown~\cite{Davighi:2019rcd} that a SM$\times U(1)_X$
gauge theory suffers from no global gauge anomalies provided only that the $SU(2)_L$
anomaly~\cite{Witten:1982fp} cancels, as it does here. 
}. 
Our model here will not be UV complete, since new physics above the TeV scale is required to generate the light Yukawa couplings (once integrated out and matched onto our SM$\times U(1)_X$ model). Nevertheless, it is wise to cancel gauge anomalies in the SM$\times U(1)_X$ model, as we do here; if not, there must be heavy chiral fermions to cancel anomalies, whose masses are at least tied to the TeV-scale $U(1)_X$ breaking.
Many viable anomaly-free $U(1)_X$ charge assignments have been investigated: for example muon minus tau lepton
number~\cite{Altmannshofer:2014cfa,Crivellin:2015mga,Crivellin:2015lwa,Crivellin:2015era,Altmannshofer:2015mqa,Davighi:2020qqa},
third family baryon number minus second family lepton
number~\cite{Alonso:2017uky,Bonilla:2017lsq,Allanach:2020kss}, third family
hypercharge~\cite{Allanach:2018lvl,Davighi:2019jwf,Allanach:2019iiy} or other
assignments~\cite{Sierra:2015fma,Celis:2015ara,Greljo:2015mma,Falkowski:2015zwa,Chiang:2016qov,Boucenna:2016wpr,Boucenna:2016qad,Ko:2017lzd,Alonso:2017bff,Tang:2017gkz,Bhatia:2017tgo,Fuyuto:2017sys,Bian:2017xzg,King:2018fcg,Duan:2018akc,Kang:2019vng,Calibbi:2019lvs,Altmannshofer:2019xda,Capdevila:2020rrl,Davighi:2021oel}. 

In Ref.~\cite{Allanach:2018lvl}, the Third Family Hypercharge
($Y_3$) model was presented and shown to fit the $b\rightarrow s \ell \ell$
data. Here, the $X$ charges of all fermions are zero except for the third
family, which has $X$ charge equal to its hypercharge. The model predicts
$Z-Z^\prime$ mixing because the Higgs doublet is necessarily charged under
$U(1)_X$ to allow a renormalisable top Yukawa coupling, which would
otherwise be forbidden by the $U(1)_X$ gauge symmetry.
It was noted that this
will change the SM prediction for $M_W$~\cite{Davighi:2020nhv} and in
Ref.~\cite{Allanach:2021kzj} it was shown that the model could simultaneously
fit the $b\rightarrow s \ell \ell$ anomalies and EWPOs: indeed, that the
SM prediction for $M_W$ was improved, since it was some $2 \sigma$ too low in
the SM (compared to the measurement at the time) but receiving a positive
correction from the $Y_3$ model. 

We will show below that including the new CDF II
measurement of $M_W$ as in (\ref{avMW}), the $Y_3$ model is somewhat disfavoured with a global
$p-$value of .02. Our purpose in this paper is then to build a generalisation
of the $Y_3$ model, which is similar in construction and as simple, but which
provides a simultaneously acceptable fit to both the $M_W$ anomaly and
$b\rightarrow s \ell \ell$ anomalies. To this end, in \S\ref{sec:models}, we
shall propose a
generalisation of the $Y_3$ charge assignments, which is rendered anomaly-free 
simply by the inclusion of right-handed neutrinos. There are two classes of
charge assignment, depending upon whether one permutes the \emph{right-handed}
third family field's charged lepton charge with that of the second family
(i.e.\ $X_{e_3} \leftrightarrow X_{e_2}$) or not.
In \S\ref{sec:SMEFT}, we
calculate the SM effective field theory (SMEFT) coefficients that arise from
integrating out the heavy $Z^\prime$ boson, and then matching to the SMEFT at
tree level. 
This allows us to encode our
model in a suitable form for input into \smelli\ and use its calculation of
EWPOs, LFU observables and the `quarks' category of observables. We discuss
the 
phenomenology of the models in \S\ref{sec:pheno}, in particular the effect on
$M_W$ and on $b\rightarrow s \ell \ell$ coefficients in the weak effective
theory. The case where one permutes
$X_{e_3} \leftrightarrow X_{e_2}$
corresponds to $P_X=P_L+P_R$, 
whereas in the case where one does not, one
obtains a certain linear combination of $P_L$ and $P_R$, depending upon $s$
and $t$. In both cases, there is a purely axial coupling to electrons.
We then present global fits to this latter family of model. The
inclusion of the 2022 CDF II $M_W$ 
measurement selects a subset of models which provide acceptable fits to the
collective data. We conclude in \S\ref{sec:conc}. 

\section{The models} \label{sec:models}


We consider a class of $\ZP$ models based on extending the SM gauge symmetry by a $U(1)_X$ factor, where
\be
X = s Y_3 + t (B-L)_3\, , \qquad s \in \mathbb{N},\,  t \in \mathbb{Z}, \label{sandt}
\ee
$Y_3$ is third family hypercharge and $(B-L)_3$ is third family baryon
number minus lepton number.
Our conventions for the representations of the non-gauge fields in the model
are shown in Table~\ref{tab:fields}.
\begin{table}
\begin{equation*}
\begin{array}{|c|ccc|ccc|ccc|ccc|ccc|ccc|c|c|}
\hline
& q_1^\prime & q_2^\prime & q_3^\prime & u_1^\prime& u_2^\prime & u_3^\prime & d_1^\prime & d_2^\prime & d_3^\prime & \ell_1^\prime & \ell_2^\prime &
\ell_3^\prime &e_1^\prime & e_2^\prime &e_3^\prime & \nu_1^\prime &\nu_2^\prime &\nu_3^\prime & H & \theta \\ \hline
 SU(3) && {\bf 3} &  &  & {\bf 3} &  & & {\bf 3} &  & & {\bf 1} &  & & {\bf 1}
 & &  & {\bf 1} & & {\bf 1} & {\bf 1}\\
 SU(2) & & {\bf 2} & &  & {\bf 1} &  & & {\bf 1} & & & {\bf 2} & & & {\bf 1} &
 & & {\bf 1}  & & {\bf 2} & {\bf 1} \\
 U(1)_Y & & 1 &  &  & 4 &  & & -2 & & & -3 & & & -6 & &  & 0 & & 3 & 0\\ 
\hline
 U(1)_{Y_3} & 0 & 0 & 1& 0 & 0 & 4 & 0 & 0 & -2 & 0 & 0 & -3 & 0 & 0 & -6 & 0
 & 0 & 0 & 3 & \ast\\
 U(1)_{(B-L)_3} & 0 & 0 & 1 & 0 & 0 & 1 & 0 & 0 & 1 & 0 & 0 & -3 & 0 & 0 & -3
 & 0 & 0 & -3 & 0 & \ast\\
\hline
\end{array}
\end{equation*}
\caption{Representations of fields under the SM gauge factors, which
  are family universal, together with the family non-universal $Y_3$ and
  $(B-L)_3$ symmetries on which our $\ZP$ model is based. We use the minimal
  integer normalisation for the charges under each $U(1)$ factor. 
Note that the
  permutations $\ell_2 \leftrightarrow \ell_3$ (and in some cases $e_2
  \leftrightarrow e_3$) are made \emph{after} the assignments shown.
  All fields are Weyl fermions except for the complex scalar Higgs doublet $H$
  and the complex scalar flavon $\theta$. 
  $\ast$ denotes that the charge is a non-zero number whose value does not
  change any of the discussion or results of this paper.
 \label{tab:fields}
}
\end{table}
The $X$ charge assignment in 
(\ref{sandt}) is the most general anomaly-free $U(1)_X$
extension\footnote{We ignore possible discrete quotients of the gauge group because they are not
  relevant to our discussion.} that couples only to a single family of SM
fermions~\cite{Allanach:2018vjg}, including a right-handed neutrino. While we have parameterised this family of
$U(1)_X$ models by two integers $s$ and $t$, it is only the rational parameter
$t/s \in \mathbb{Q}$ that is relevant for phenomenology\footnote{This is
  because all charges may be scaled by an overall factor, so long as the
  $U(1)_X$ gauge coupling $g_X$ is scaled by one over this factor, with no change to
  the physics.}.
This family of $U(1)_X$ extensions of the SM are known to have semi-simple gauge completions without needing any extra chiral fermions, as was shown in Ref.~\cite{Davighi:2022dyq}.

The $\ZP$ model is designed to explain the $b\to
s\ell\ell$ anomalies. Global fits strongly favour a lepton flavour non-universal
coupling of the $\ZP$ to left-handed muons, at least, as described in \S\ref{sec:int}. We thus permute the non-zero left-handed lepton charge from the
third to the second family, as in the original $Y_3$ model
of~\cite{Allanach:2018lvl}. Regarding right-handed leptons, 
acceptable fits can be obtained with or without permuting the non-zero
right-handed lepton charge to the second family. 

In summary, the charges of the SM fermions, together with the SM Higgs doublet
$H=(H^+,H^0)^T$ and an extra $U(1)_X$ symmetry-breaking scalar $\theta$ which
is a SM singlet, are 
\begin{align} 
X_{q_3} &= s+t\,, \qquad &&X_{\ell_2} = -3s-3t\,, \label{impQ}\\
X_{u_3} &= 4s+t\,, \qquad &&X_{e_n} = -6s-3t\,,\,\,\,\, n=2\text{~or~}3\,,\\
X_{d_3} &= -2s+t\,, \qquad &&X_{\nu_n} = -3t\,, \\
X_{H}&=3s\,, \qquad &&X_{\theta} \label{eq:charges},
\end{align}
where $X_\theta \neq 0$, with all other $U(1)_X$ charges being zero. 
The original $Y_3$ model of Ref.~\cite{Allanach:2018lvl}, which is a template
for the family of models we consider here, can be recovered by setting $n=3$ and $(s,t)=(1,0)$, thus decoupling the right-handed neutrino.

Like the $Y_3$ model~\cite{Allanach:2018lvl}, the family of models we consider
here allow, on the quark side, only third family Yukawa couplings to the Higgs
at the renormalisable level, $\mathcal{L} \supset y_t \overline{q}_3^\prime
H^c u_3^\prime  + y_b \overline{q}_3^\prime H d_3^\prime$. Thus, to zeroth
order, gauging the anomaly-free chiral third-family symmetry
(\ref{eq:charges}) postdicts a heavy third family and small quark mixing
angles, as observed. The light quark Yukawa couplings, responsible for the
masses of the first and second generation and for the quark mixing angles,
must come from higher-dimensional operators. Such operators can come from
integrating out heavier fermionic representations that are vector-like under to
the gauge group. These are details of the UV theory which it would
be premature to specify (although see
Ref.~\cite{Allanach:2021kzj} for further comments and ideas). 

On the other hand, the lepton sector is not so natural from the Yukawa perspective, due to the fact that we permute the non-zero $X_{\ell_i}$ charge into the second family. At face value, the charge assignment (\ref{eq:charges}) with $n=3$ allows only for a renormalisable off-diagonal Yukawa coupling $\sim \overline{\ell}_2^\prime H e_3^\prime$, which must be highly suppressed in order to explain the relative heaviness of the tau and the non-observation of $\mu-\tau$ lepton flavour violation. This is easily explained with a little more model-building; for example, additionally gauging an anomaly-free lepton-flavoured $U(1)$ symmetry could ban this off-diagonal Yukawa coupling.\footnote{Another route is to look for anomaly-free deformations of (\ref{eq:charges}) in which two families of leptons are charged, generalising~\cite{Allanach:2019iiy}, which allow either no renormalisable lepton Yukawa couplings, or at most the $y_\tau\overline{\ell}_3^\prime H e_3^\prime$ coupling.
}


In what follows we use a convention in which the covariant derivative acting on a field $f$ is
\begin{equation}
D_\mu = \partial_\mu - i g \frac{\sigma^a}{2} P_L W_\mu^a - i g^\prime Y_f B_\mu - i g_X X_f X_\mu\, ,
\end{equation}
where $X_\mu$ is the gauge field for $U(1)_X$ and $g_X$ is its gauge coupling.
We assume that any kinetic mixing between $U(1)_X$ and $U(1)_Y$ gauge
fields is negligible, leaving a study of such effects to future work.
The fermion couplings to the gauge fields, in the gauge eigenbasis (indicated by `primes'), are
\begin{equation} \label{eq:psi_coup}
\mathcal{L}_{\psi} = g_X X_\mu \sum_{\psi} X_{\psi} \bar{\psi}^\prime \gamma^\mu \psi^\prime\, ,
\end{equation}
where the sum on $\psi$ runs over all SM Weyl fermions and the charges $X_\psi$ are those in (\ref{eq:charges}).

\subsection*{Symmetry breaking and {\boldmath$Z-\ZP$} mixing}

The $U(1)_X$ symmetry is broken predominantly by the SM singlet scalar field
$\theta$, which acquires a vacuum expectation value $\langle \theta \rangle =
v_X/\sqrt{2}$, where $v_X \gg v$ is of order the TeV scale. This implies that
the mass of the $Z^\prime$ field $M_{\ZP} \approx g_X X_\theta v_X$. 
However, because the SM Higgs is also charged under $U(1)_X$, there is mass
mixing 
between the $Z$ boson and the $\ZP$ gauge boson. These physical gauge bosons
are linear combinations of the SM combination\footnote{We define $c_x:=\cos
  x$, $s_x:=\sin x$ throughout this paper, for various quantities $x$.}
$Z^0_\mu = c_w W^3_\mu - s_w B_\mu$ (where $\theta_w = \text{tan}^{-1}(g'/g)$ is the Weinberg angle) and the $X_\mu$ field {\em viz.}
\begin{equation}
\begin{pmatrix} Z_\mu \\ Z^\prime_\mu \end{pmatrix} = 
\begin{pmatrix} c_z & s_z \\ -s_z & c_z \end{pmatrix}
\begin{pmatrix} Z^0_\mu \\ X_\mu \end{pmatrix},
\end{equation}
where the mixing angle $\alpha_z$ is determined by, at tree-level,
\begin{equation} \label{eq:angle}
\sin\alpha_z = \frac{X_H}{2X_\theta^2} \frac{g/c_w}{g_X} \frac{v^2}{v_X^2}=
\frac{2X_H g_X}{g/c_w}\frac{M_Z^2}{M_{\ZP}^2} \left[ 1 +
\mathcal{O}\left(\frac{M_Z^2}{M_{\ZP}^2}\right) \right] ,
\end{equation}
where
\begin{equation}M_Z^2 =  v^2 g^2/4c_w^2 \left[ 1 +
    \mathcal{O}\left(\frac{M_Z^2}{M_{\ZP}^2}\right)\right]. \label{mz2}
\end{equation}
At a constant value of $M_Z$ (we shall take $M_Z$ to be an experimental
input), 
the ${\mathcal O}(M_Z^2/M_{\ZP}^2)$ 
correction in (\ref{mz2}) 
translates to an upward shift in the SM prediction  
for $M_W$, as we shall describe in \S\ref{sec:observables}. 

\subsection*{Fermion mixing matrices}

Following Refs.~\cite{Allanach:2018lvl,Allanach:2021kzj}, we assume a simple
ansatz for the 3-by-3 unitary fermion mixing matrices describing the change
from the gauge eigenbasis to the (unprimed) mass eigenbasis of the fermionic fields.
The purpose of this ansatz is to characterise the main physical
flavour characteristics of the model
without introducing large flavour-changing neutral currents that would be
subject to strong experimental constraints. We may think of the ansatz as a
limit to expand around: the fairly strong assumptions might be motivated by
further model-building involving additional symmetries or dynamics, but we
leave such considerations to one side, for now.

For left-handed down-type quarks, we parameterise the mixing matrix
as
\begin{equation}
  V_{d_L}=
\left(\begin{array}{ccc}
1 & 0 & 0  \\
0 & \cos \theta_{sb} & \sin \theta_{sb} \\
0 & -\sin \theta_{sb} & \cos \theta_{sb} \\
\end{array} \right).
\end{equation}
For simplicity, and to avoid unwanted large contributions to $\Delta F = 2$ processes and charged lepton flavour violating processes, we 
choose $V_{d_R}=1$, $V_{u_R}=1$ and $V_{e_L}=1$. 
Finally, $V_{u_L} = V_{d_L}V^\dagger$ and $V_{\nu_L} = V_{e_L}U^\dagger$ are
fixed by the CKM matrix $V$ and the PMNS matrix $U$, respectively.
We use the
central values of extracted angles and phases from the Particle Data
Group~\cite{ParticleDataGroup:2020ssz}. 

Using these rotation matrices, the couplings of the $\ZP$ boson to the physical fermion states can be obtained from (\ref{eq:psi_coup}). 
For the left-handed quarks, we work in a `down-aligned' basis in which the
(unprimed) quark doublets are $q_i = (V^\dagger_{ij} u_{L,j},\ ~d_{L,i})$.
For the left-handed leptons, we work in a basis in which the left-handed charged leptons $e_{L_i}$ align with the mass eigenstates, and thus the (unprimed) lepton doublets are $\ell_i = (U^\dag_{ij} \nu_{L,j},\ ~e_{L, i})$.
Since $V_{e_L} = V_{e_R}=1$ the charged lepton couplings remain diagonal. The
down quark couplings are mixed away from the diagonal, however, as is indeed necessary to obtain a quark
flavour-violating coupling of the $\ZP$ to $b\bar{s}$ and $\bar{s}b$. We have: 
\begin{align}
\mathcal{L}_{\psi} \supset g_X X_\mu &\big( X_{q_3}\Lambda^{d_L}_{ij} \overline{q}_i \gamma^\mu q_j + X_{u_3} \overline{u}_3 \gamma^\mu u_3 + X_{d_3} \overline{d}_3 \gamma^\mu d_3  \\
&+ X_{\ell_2} \overline{\ell}_2 \gamma^\mu \ell_2 + X_{e_n} \overline{e}_n \gamma^\mu e_n + X_{\nu_n} \overline{\nu}_n \gamma^\mu \nu_n  \big) \nonumber \, ,
\end{align}
where $\Lambda^{d_L}_{ij} := V_{d_L}^\dagger \text{diag}(0,0,1) V_{d_L}$.

\section{SMEFT matching} \label{sec:SMEFT}

Integrating out the heavy $X$ boson and matching onto the SM Effective Field
Theory (SMEFT) at a scale $M_X:=g_X X_\theta v_X$, we obtain the Wilson coefficients (WCs) for
dimension-6 SMEFT operators written in Table~\ref{tab:WCs}. The WCs $\{C_i\}$ have
units of $[\text{mass}]^{-2}$ and are written in the Warsaw
basis~\cite{Grzadkowski:2010es}, a basis of a set of independent baryon-number
conserving operators. By performing the matching between our models and
the SMEFT, we obtain the set of WCs $\{C_i\}$ at the scale $M_X$, which can then
be used to calculate predictions for observables.
\begin{table}
  \begin{center}
     \begin{tabular}{|c|c||c|c|} \hline
       WC & Value & WC & Value \\ \hline
       $C_{ll}^{2222}$&  $-\frac{1}{2} X_{\ell_2}^2$      &
       $ (C_{lq}^{(1)})^{22ij}$& $-X_{q_3}X_{\ell_2} \Lambda_{ij}^{d_L}$\\
       $(C_{qq}^{(1)})^{ijkl}$&$ X_{q_3}^2{\Lambda_{ij}^{d_L}} {\Lambda_{kl}^{d_L}}\frac{\delta_{ik}\delta_{jl}-2}{2}$ &
       $C_{ee}^{nnnn}$& $-\frac{1}{2} X_{e_n}^2$\\     
       $C_{uu}^{3333}$& $-\frac{1}{2} X_{u_3}^2$ &     
       $C_{dd}^{3333}$& $-\frac{1}{2} X_{d_3}^2$ \\     
       $C_{eu}^{nn33}$& $-X_{e_n}X_{u_3}$ &     
       $C_{ed}^{nn33}$& $-X_{e_3}X_{d_3}$ \\     
       $(C_{ud}^{(1)})^{3333}$& $-X_{u_3}X_{d_3}$&
       $C_{le}^{22nn}$& $-X_{\ell_2}X_{e_n}$\\     
       $C_{lu}^{2233}$& $-X_{\ell_2}X_{u_3}$ &     
       $C_{ld}^{2233}$& $-X_{\ell_2}X_{d_3}$ \\     
       $C_{qe}^{ijnn}$& $-X_{q_3}X_{e_n} \Lambda_{ij}^{d_L}$ &     
       $(C_{qu}^{(1)})^{ij33}$& $-X_{q_3}X_{u_3} \Lambda_{ij}^{d_L}$\\     
       $(C_{qd}^{(1)})^{ij33}$& $-X_{q_3}X_{d_3} \Lambda_{ij}^{d_L}$& 
       $(C_{\phi l}^{(1)})^{22}$& $-X_H X_{\ell_2}$\\
       $(C_{\phi q}^{(1)})^{ij}$& $-X_H X_{q_3}$ &
       $C_{\phi e}^{nn}$& $-X_H X_{e_n}$\\
       $C_{\phi u}^{33}$& $-X_H X_{u_3}$&
       $C_{\phi d}^{33}$& $-X_H X_{d_3}$\\       
       $C_{\phi D}$& $-2 X_H^2$&
       $C_{\phi \Box}$&$-\frac{1}{2} X_H^2$ \\       
 \hline  \end{tabular}
  \end{center}
  \caption{\label{tab:WCs} Non-zero dimension-6 SMEFT Wilson coefficients in the Warsaw basis, obtained by integrating out the heavy $X$ boson at scale
    $M_{X}$. We write the coefficients as functions of the charges $X_f$, which are explicitly parameterised in (\ref{eq:charges}). The integer $n=2$ or $n=3$ corresponds to two variations of the model, as explained in \S\ref{sec:models}. All Wilson coefficients are in units of $g_X^2 / M_X^2$.  }
\end{table}

\section{Phenomenology \label{sec:pheno}}

Starting from the SMEFT matching of \S\ref{sec:SMEFT}, we will use the
\smelli\ program to evaluate the likelihood of the model given hundreds of
observables in the electroweak and flavour sectors. Before we do so, however,
we think it important to highlight the most important observables that are
sensitive to our model, and how these depend upon the SMEFT coefficients and
hence upon the $X_f$ charges. 

\subsection{Important observables} \label{sec:observables}

\subsubsection*{Electroweak}

In light of the recent CDF II measurement~\cite{CDF:2022hxs}, the $M_W$
prediction is
deserving of special attention. We can parameterise its deviation from the SM
prediction in our $\ZP$ model via the parameter
\be \label{eq:rho}
\rho_0 := \frac{M_W^2}{M_Z^2 \hat{c}_Z^2 \hat{\rho}},
\ee
where the parameter $\hat{\rho}=1.01019\pm 0.00009$ includes the
custodial-violating top contributions to the gauge boson masses (see Ref.~\cite{ParticleDataGroup:2020ssz}).
In~(\ref{eq:rho}), we use the conventional notation $\hat{c}_Z^2 := \cos^2 \hat{\theta}_w(M_Z)=\frac{\hat{g}^2(M_Z)}{\hat{g}^2(M_Z)+\hat{g}'^2(M_Z)}$ to denote the cosine squared of the renormalised Weinberg angle in the $\overline{\text{MS}}$ scheme.
The $\rho_0$ parameter is defined so as to equal precisely 1 in the SM using the $\overline{\text{MS}}$ scheme. 
Its deviation from unity in our heavy $\ZP$ model, which recall is due to $Z-\ZP$ mixing, is
\begin{equation}
\rho_0 \approx 1 + \frac{4X_H^2 g_X^2}{g^2+g^{\prime 2}} \frac{M_Z^2}{M_{\ZP}^2}\, .
\end{equation}
Importantly, $\rho_0$ unavoidably shifts {\em
  upwards}~\cite{Allanach:2021kzj,Alguero:2022est}, easing the tension due to
the CDF II $M_W$ measurement irrespective of the sign
of the Higgs charge $X_H$.

At the level of the SMEFT, this shift is captured by the Wilson coefficients $C_{\phi D}$ and $C_{\phi \Box}$, which are (as in Table~\ref{tab:WCs})
\begin{equation}
C_{\phi D} = 4 C_{\phi \Box} = -2X_H^2 \frac{g_X^2}{M_X^2} = -18s^2 \frac{g_X^2}{M_X^2}\, .
\end{equation}
Global fits of the SMEFT to electroweak data that turn on only these two operators
have been found to give a good fit to electroweak data in light of the CDF
measurement ({\em e.g.}~\cite{Bagnaschi:2022whn}). 

The status of the
electroweak fit is of course much more delicate in our model for the $b\to
s\ell\ell$ anomalies, because we have a plethora of new physics effects in $Z$
pole observables due to modified $Z$ couplings to fermions. Important among these are modified $Z$ couplings to leptons, especially muons, 
and the forward-backward asymmetry variable $A^{FB}_b$ for $Z$ decays to $b\bar{b}$, because these observables already exhibit small tensions with the SM and because they receive large-ish corrections in our model.
These effects are strictly correlated to the shifts in $C_{\phi D}$ and
$C_{\phi \Box}$, since they
arise from the $Z-\ZP$ mixing.

\subsubsection*{{\boldmath$b\to s\ell\ell$}}

Regarding the $b\to s\ell\ell$ anomalies and related observables, the new physics effects can be parameterised by contributions to Wilson coefficients in the Weak Effective Theory (WET). The WET Hamiltonian is, using a conventional normalisation,
\be
\mathcal{H}_{\text{EFT}} = - \frac{4G_F}{\sqrt{2}} V_{tb} V_{ts}^\ast \sum_i (C_i^{\text{SM}} + C_i) \mathcal{O}_i\, ,
\ee
where we emphasise that in the present paper, $C_i$ denotes the \emph{beyond} the SM (BSM) contribution to the Wilson coefficient. 
When coupling to 
$\bar{b}s$/$\bar{s}b$ currents, the $Z^\prime$ only couples to the left-handed
component and so
there are contributions to the dimension-6 semi-leptonic operators
\begin{align}
\mathcal{O}_9^{\ell\ell} &=  \frac{e^2}{16\pi^2} (\bar{s} \gamma_\mu P_L b)(\bar{l}\gamma^\mu \ell)  , \qquad \ell \in \{e,\mu\}\, ,\nonumber \\
\mathcal{O}_{10}^{\ell\ell} &= \frac{e^2}{16\pi^2} (\bar{s} \gamma_\mu P_L b)(\bar{l}\gamma^\mu\gamma_5 \ell) \, .
\end{align}

Because of the $Z-\ZP$ mixing, the $Z$ boson picks up a small quark
flavour-violating coupling to $\bar{b}s$, {\em viz.} $\mathcal{L} \supset
g_Z^{sb} Z_\mu \bar{b}_L \gamma^\mu s_L+H.c.$, where the coupling $g_Z^{sb}$
is proportional to the $Z-\ZP$ mixing angle~(\ref{eq:angle}).
Specifically,
\begin{equation}
  g_Z^{sb} = X_{q_3} X_H \sin 2\theta_{sb} \frac{g_X^2}{g/c_w}
  \frac{M_Z^2}{M_Z'^2}.
  \end{equation}
There are thus BSM contributions to the 4-fermion operators
$\mathcal{O}_{9,10}^{\ell\ell}$ from both $Z$ exchange and from $\ZP$ exchange, and these are of the same order. The $Z$ contributions are lepton flavour universal (LFU), whereas the $\ZP$ contributions introduce lepton flavour universality violation (LFUV). 

Accounting for both contributions\footnote{Of course these contributions to the WET coefficients can also be derived at the level of the SMEFT matching. From this perspective, the SMEFT-WET matching formulae are~\cite{Aebischer:2015fzz,Jenkins:2017jig}:
\begin{align} \label{eq:WCs-from-matching}
C_9^{ij} &= \frac{4\pi^2}{e^2} \frac{v^2}{V_{tb} V_{ts}^\ast} \left[-\frac{1}{k}\left( C_{\phi q}^{(1)\, 23} + C_{\phi q}^{(3)\, 23}\right)  + C_{\ell q}^{(1)\, ij23} +   C_{\ell q}^{(3)\, ij23} +  C_{qe}^{23ij} \right] \\
C_{10}^{ij} &= \frac{4\pi^2}{e^2} \frac{v^2}{V_{tb} V_{ts}^\ast} \left[ C_{\phi q}^{(1)\, 23} + C_{\phi q}^{(3)\, 23} - C_{\ell q}^{(1)\, ij23} -   C_{\ell q}^{(3)\, ij23} +  C_{qe}^{23ij}\right]\, . \nonumber
\end{align}
Substituting in the SMEFT Wilson coefficients from Table~\ref{tab:WCs} reproduces (\ref{eq:C9etc}).
} 
we have
\begin{align} \label{eq:C9etc}
 C_{9}^{\mu\mu} &= -N(X_{\ell_2}+X_{e_2}) +  C_9^U, \quad &&\text{where~} C_9^U = NX_H/k \approx 0, \nonumber \\
 C_{10}^{\mu\mu} &= -N(-X_{\ell_2}+X_{e_2}) + C_{10}^U, \quad &&\text{where~} C_{10}^U = -NX_H , \nonumber \\
 C_{9}^{ee} &=  C_9^U \approx 0, \nonumber \\
 C_{10}^{ee} &=  C_{10}^U,
\end{align}
where recall that $k=1/(1-4\sin^2 \theta_w) \approx 9.23$. We have defined the common pre-factor
\be \label{eq:N}
N=\frac{\sqrt{2}\pi^2}{e^2 G_F} \frac{\sin 2\theta_{sb}}{V_{tb} V_{ts}^\ast}\frac{g_X^2 X_{q_3}}{M_{\ZP}^2},
\ee
where $e$ is the electromagnetic gauge coupling. 
The $Z$-induced LFU pieces in (\ref{eq:C9etc}) clearly vanish in the limit
$X_H\to 0$, in which there is no $Z-\ZP$ mixing. The
LFU contributions to the $b \to s \ell\ell$ observables are therefore
correlated to their effects on EWPOs, including the shift in $M_W$. This
correlation was explored, prior to the updated CDF II measurement, in
Ref.~\cite{Alguero:2022est}. 

Substituting in the parametrisation (\ref{impQ}--\ref{eq:charges}) of charges in our models, and dropping the $1/k$ suppressed $C_9^U$ contributions, we have
\begin{align}
n=3: \qquad 
 C_{9}^{\mu\mu} &\approx{3N} (-s-t) , \nonumber \\ 
 C_{10}^{\mu\mu} &= {3N}(2s+t)  , \nonumber \\
 C_{10}^{ee} &=  {3N}s.
\end{align}
This is equivalent to writing the chirality of the coupling to muons in terms
of left and right projection operators as in \S\ref{sec:int}, i.e.\ via an effective operator $\propto (\bar{s} \gamma_\mu P_L b)(\bar{\mu} \gamma^\mu P_X \mu)$. For a general $s$ and $t$, this projection operator is
\be
P_X = P_L - \frac{s}{3s+2t}P_R\,  \qquad (n=3)\, .
\ee
For the variation of the model with $n=2$, we have
\begin{align}
n=2: \qquad 
 C_{9}^{\mu\mu} &\approx 3N(-3s-2t) , \nonumber \\ 
 C_{10}^{\mu\mu} &= 0 , \nonumber \\
 C_{10}^{ee} &= 3Ns \, ,
\end{align}
thus $P_X \approx P_L+P_R$, a vectorial coupling of the $\ZP$ to $\mu^+\mu^-$.

\subsection{Global fits} \label{sec:fits}

We wrote a computer program (included in the ancillary information attached
to the {\tt arXiv} preprint of this paper) that uses \smelli\ to calculate the
likelihoods of 277 observables for our model by first calculating the WCs in
Table~\ref{tab:WCs} at scale $M_X$. \smelli\ then renormalises the SMEFT
operators down to the scale $M_Z$, where it calculates the EWPOs. It then matches at
tree-level to the weak effective theory, and renormalises the resulting WCs down to the scale
of the mass of the bottom quark, where the various observables pertinent to
$B-$meson decays are calculated. 

For a given $s$ and $t$
the likelihoods are a function of
two effective parameters: $\alpha:=g_X \times \text{3 TeV}/M_X$ and
the quark mixing angle $\theta_{sb}$. Throughout, we shall
illustrate with $M_X=3$ TeV as an example (similar third-family type models
with $M_X\geq 1$ TeV
were not ruled out by search constraints in the parameter region where
they fit the 
$b\rightarrow s \ell \ell$ anomalies~\cite{Allanach:2021bbd,Allanach:2021gmj}
and so we expect $M_X = 3$ TeV to be allowed by direct $Z^\prime$ search
constraints). 
We do not expect a significant change if
we were to change $M_X$ to a different value $M_{X^\prime}$ as long as one
scales $g_X$ by the same factor, 
since the boundary conditions upon the Wilson coefficients (shown in
Table~\ref{tab:WCs}) depend only upon the ratio $g_X/M_X$. This
approximation is good up to small loop suppressed corrections from the
renormalisation group running between $M_{X^\prime}$ and $M_{X}$, which 
induce relative multiplicative changes of ${\mathcal
  O}[\log(M_{X^\prime}/M_X)/(16 \pi^2)]$ in 
corrections to predictions of observables\footnote{Strictly speaking, we
  calculated the fit at a reference point of $M_X=1$ TeV.}.

We consider each different value of the ratio $t/s$ to constitute a different model. 
$\smelli$ then calculates $\chi^2$ via the predicted values of observables.
We 
minimise $\chi^2$ by varying $\alpha$ and $\theta_{sb}$ using the Nelder-Mead
algorithm, given a guess at a starting point. This was obtained by doing a rough
initial scan for one value of $t\neq 0$ and $s\neq 0$ and roughly estimating how
electroweak and $b-$observable $\chi^2$ values are expected to scale
with $s$ and $t$.

\begin{figure}
  \begin{center}
  \includegraphics[width=0.7\textwidth]{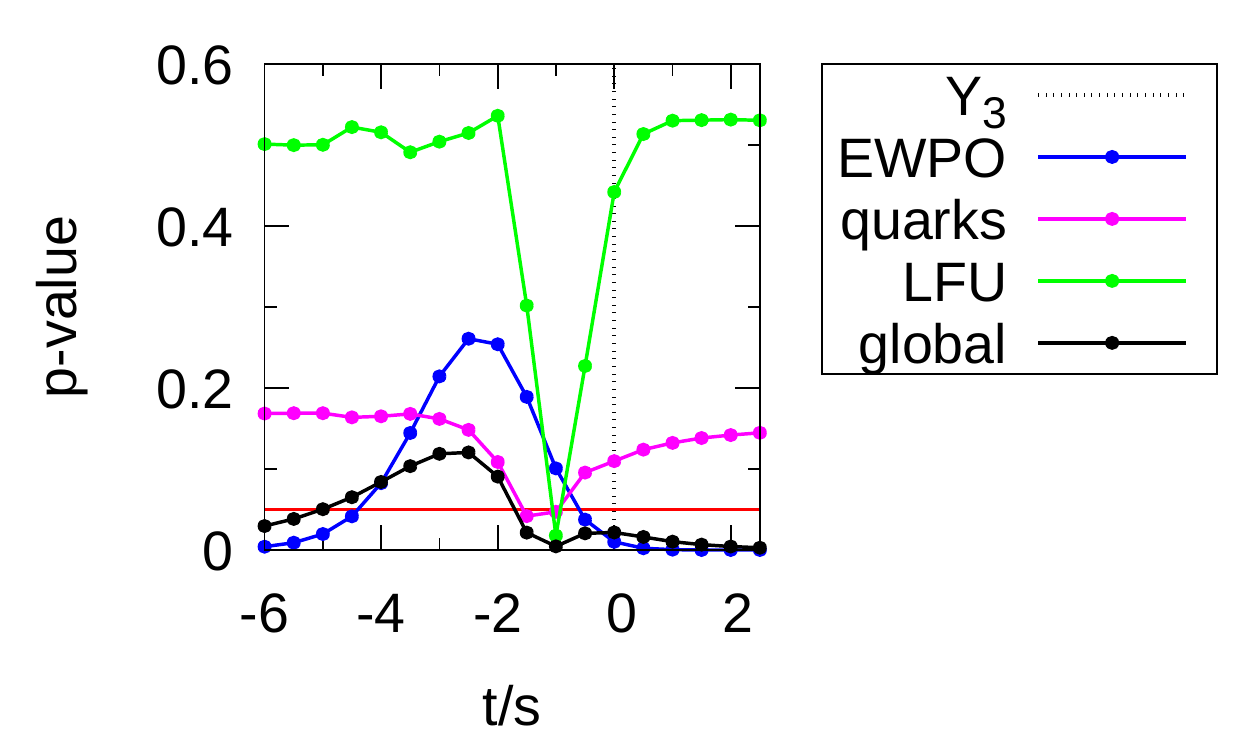}\\
  \includegraphics[width=0.7\textwidth]{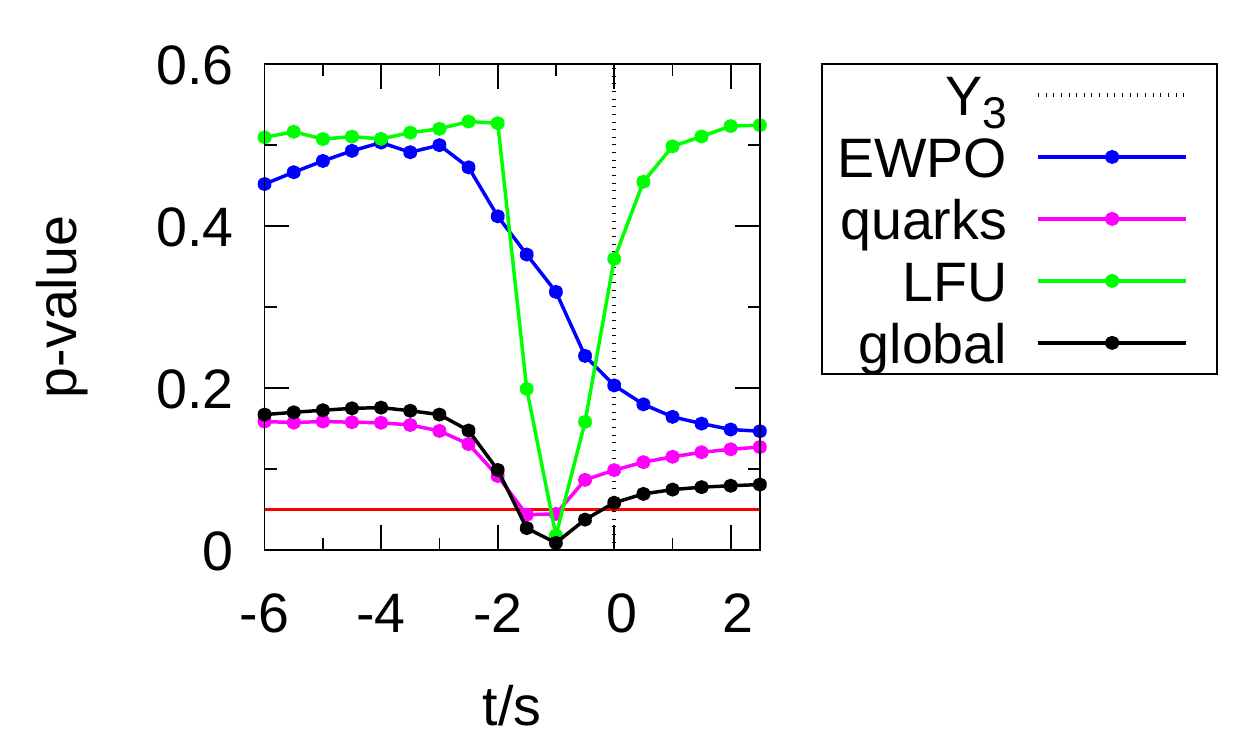}
  \caption{\label{fig:pvals} $p-$values for different models depending upon
    $t/s$, for $n=3$. The top panel includes the recent CDF measurement
    of $M_W$ as in (\ref{avMW}), whereas in the bottom panel, the default \smelli\ constraint
    (i.e.\ excluding the recent CDF measurement) on $M_W$ was used as in (\ref{smelliMW}).
        The Third Family Hypercharge Model is marked in the legend by $Y_3$.  Only the global
    $p-$value accounts for the 2 fitted parameters in the calculation of the
    number of degrees of freedom. We consider global $p-$values above the red
    line at 0.05 to be acceptable fits to the data. 
  }
  \end{center}
  \end{figure}
The result, for a given value of $t/s$, is a best-fit point,
where the fit in the EWPOs is balanced against those of the $b$ data. This
balance crucially depends upon the experimental constraint that is taken on
$M_W$, as we illustrate in the top panel of Fig.~\ref{fig:pvals}, where
we display $p-$values
using (\ref{avMW}), which
\emph{includes} the CDF II measurement. Here, we can
see that, for example, the Third Family Hypercharge Model model, denoted $Y_3$, is somewhat disfavoured:
its global $p-$value is around .02.
Models at larger values of $|t/s|$ approximate the $B_3-L_3$ model and so the
$p-$values all asymptote towards the left-hand side and the right-hand side of
the plot. The $B_3-L_3$ model has no $Z-Z^\prime$ mixing because the Higgs
doublet field is not charged under $B_3-L_3$, meaning that in the limit of
large $|t/s|$, the CDF II $M_W$ measurement strongly disfavours each model.
We see that models with $-5 < t/s < -2$
all fare well with global $p-$values above the canonical .05 bound shown by the red
line. We notice by examining the EWPO $p-$values that the $M_W$ constraint
prefers a localised value of $t/s$. The other effects that are
relevant are in the LFU and `quarks' categories: both have a valley around
$t/s=-1$. Here, at $s=-t$, one can see from (\ref{impQ}) that $X_{q_3}=X_{\ell_2}=0$, meaning that
there is no coupling at tree-level between the $Z^\prime$ and left-handed
quarks or 
left-handed muons; the $Z^\prime$ does
therefore not help with the $b\rightarrow s \ell \ell$ anomalies and we revert 
to the poor fit of the SM both for the `LFU' and for the `quarks' category of observable.
It is of interest that the $p-$value is suppressed somewhat for all $t/s$ by the
`quarks' category of observable, in which there is room for disagreement with the theoretical
uncertainty budget of the prediction.

We see a rather weak $M_W$ selection effect in the bottom panel of Fig.~\ref{fig:pvals}
for the default 
experimental $M_W$ constraint in \smelli, which amounts to
\begin{equation}
  M_W=80.3795 \pm 0.0121 \text{~GeV} \label{smelliMW}
\end{equation}
(the central value of the SM prediction according to
  \smelli\ is  $M_W=80.3509$ GeV). 
The 
\smelli\ $M_W$ constraint leads to weaker selection than the one including the
CDF II $M_W$ measurement because it has much
larger uncertainties and less need for 
a large contribution from the $Z^\prime$. We
see here that only the region $-2<t/s<0$ has a global $p-$value of \emph{less} than
.05, and this is clearly driven by the `quarks' and `LFU' categories, not by EWPOs. 
We summarise the $\chi^2$ and $p-$values for both options of experimental
$M_W$ constraint
in Table~\ref{tab:compp}. 

\begin{table}
  \begin{center}
    \begin{tabular}{|c|ccc|}\hline
      \multicolumn{4}{|c|}{including CDF II $M_W$ (\ref{avMW})} \\
      \multicolumn{4}{|c|}{$g_X\times \text{3~TeV}/M_{Z^\prime}=0.0624$,
        $\theta_{sb}=-0.0182$} \\ \hline
      category & $\chi^2$ & $n_{obs}$ & $p-$value \\ \hline
quarks & 246.2 & 224 & .15  \\
LFU FCNCs & 22.3 & 23 & .50  \\
EWPOs & 35.8 & 30 & .21  \\ \hline
global & 304.2 & 277 & .12  \\      
      \hline \hline
      \multicolumn{4}{|c|}{Default \smelli\ $M_W$ (\ref{smelliMW})} \\
      \multicolumn{4}{|c|}{$g_X\times \text{3~TeV}/M_{Z^\prime}=0.0453$, $\theta_{sb}=-0.0331$} \\      \hline
quarks & 246.4 & 224 & .14  \\
LFU FCNCs & 22.0 & 23 & .52  \\
EWPOs & 30.3 & 31 & .49  \\ \hline
global & 298.8 & 278 & .17  \\      
      \hline
    \end{tabular}
    \caption{\label{tab:compp} $p-$values for $s=2, t=-6$ for $n=3$ and the two 
    different constraints upon $M_W$. The default \smelli\ $M_W$ experimental
    constraint includes two input experimental values: one from ATLAS~\cite{ATLAS:2017rzl} and
    one combined measurement from
    CDF plus Dzero~\cite{CDF:2013dpa}, whereas we use a single combined value from Ref.~\cite{deBlas:2022hdk}
    when we include the CDF II constraint upon $M_W$. This fact explains why
    $n_{obs}$(EWPOs) differs by 1 between the two different $M_W$ constraint
    options. 
    We also display the best-fit model parameters for each option of
    experimental $M_W$ constraint.}
    \end{center}
  \end{table}

We now go on to examine the pulls at each best-fit point for $t/s=-3$. 
We define the pull for an observable with theoretical prediction $P$, measured central value $M$
and uncertainty $C$ to be
\begin{equation}
  \text{pull} = \frac{P-M}{C}, \label{pull}
\end{equation}
where $C$ does \emph{not} include correlations with other observables, but may
include theoretical uncertainties added in quadrature in some cases. In
particular, for
$M_W$, we have added an estimated uncertainty in the prediction of
5.6~MeV~\cite{deBlas:2022hdk} in quadrature.

\begin{figure}
  \includegraphics[width=0.5\textwidth]{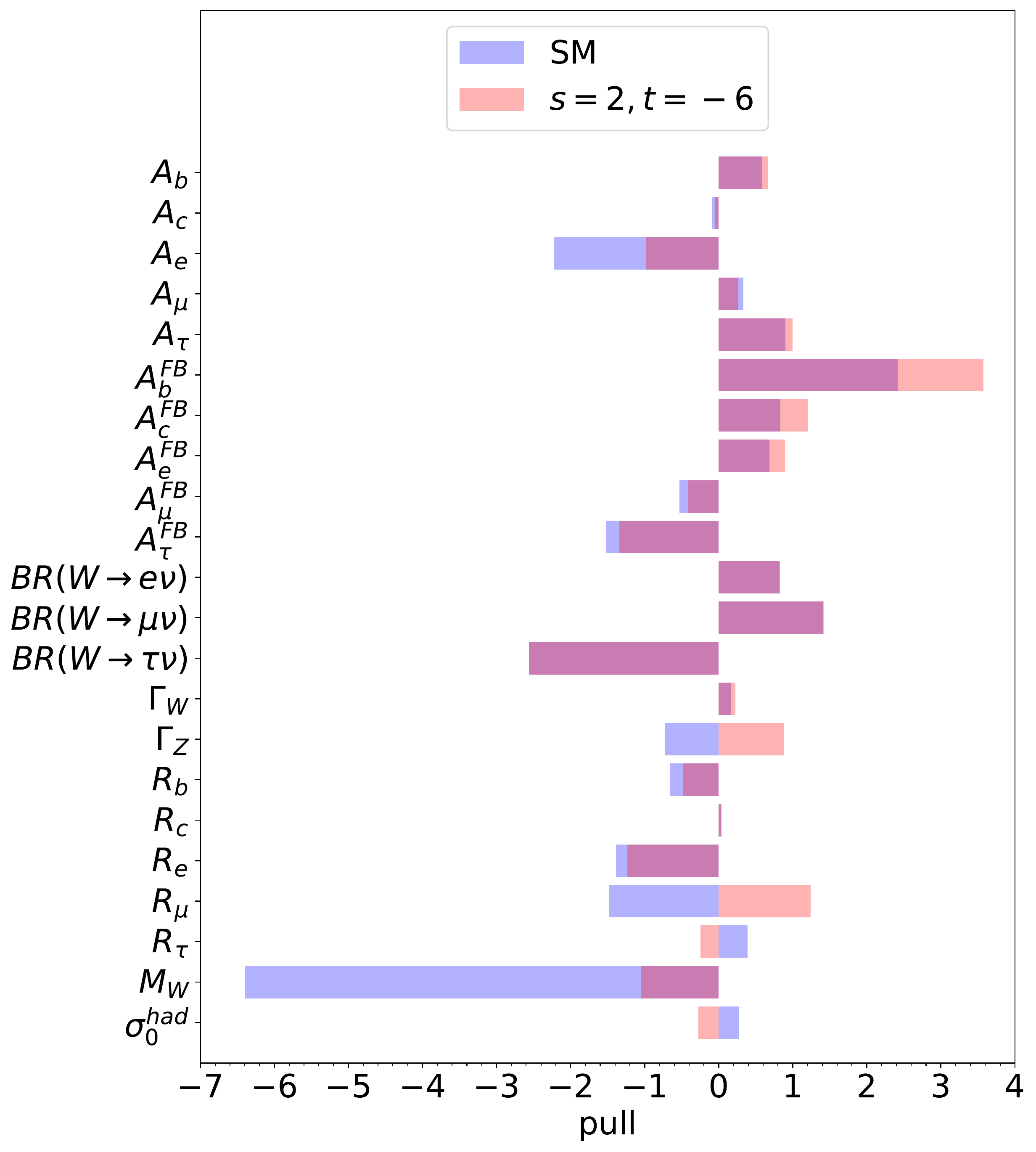}
  \includegraphics[width=0.5\textwidth]{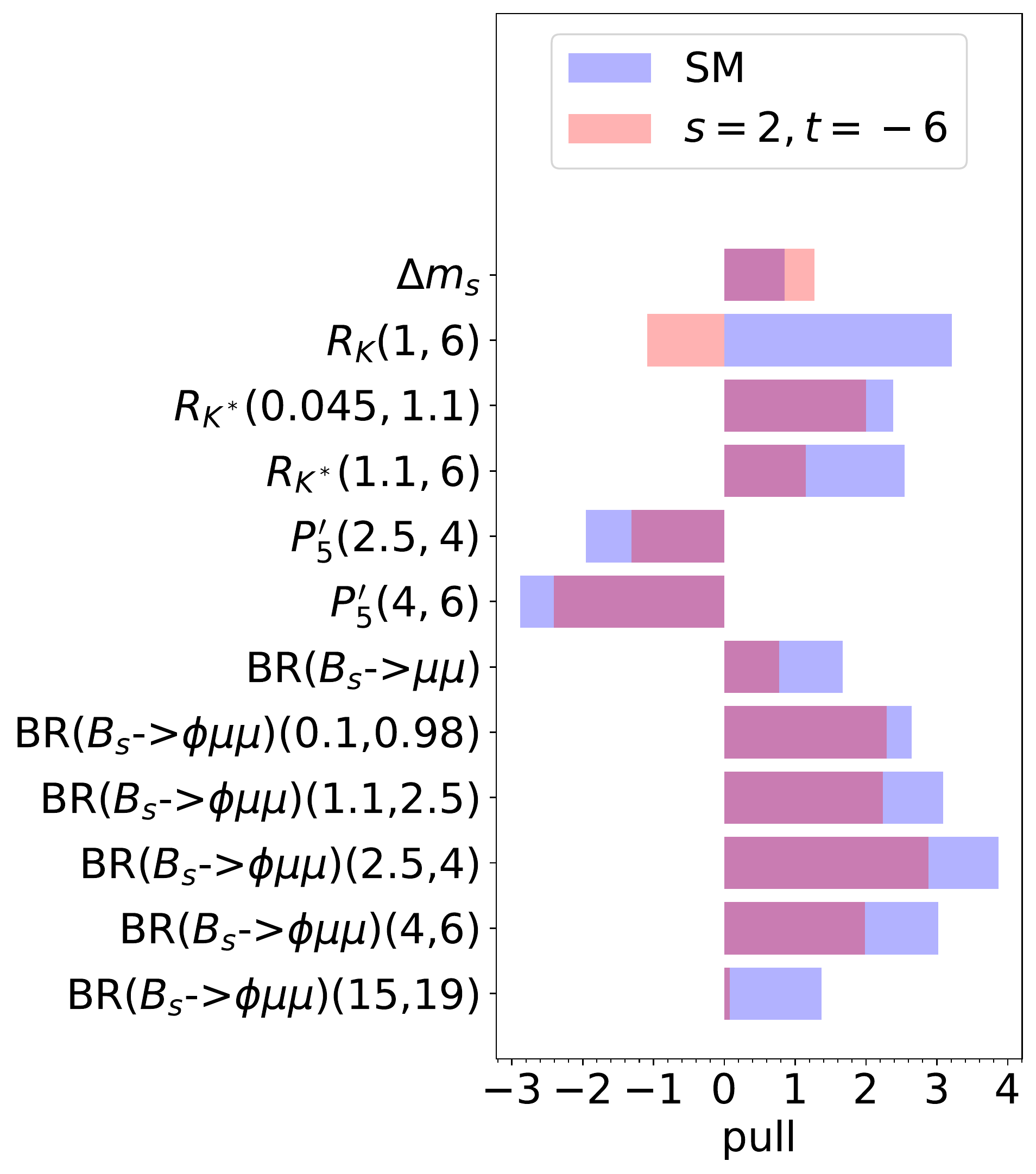}  
  \caption{\label{fig:pull_new}Pulls of interest
    including the recent CDF measurement of $M_W$ i.e.\ (\ref{avMW}) and $n=3$. 
    In the
  left-hand panel, we display the EWPOs, whereas in the right-hand panel, we
  display selected observables of interest to $b \rightarrow s l^+l^-$
  anomalies.} 
  \end{figure}
\begin{figure}
  \includegraphics[width=0.5\textwidth]{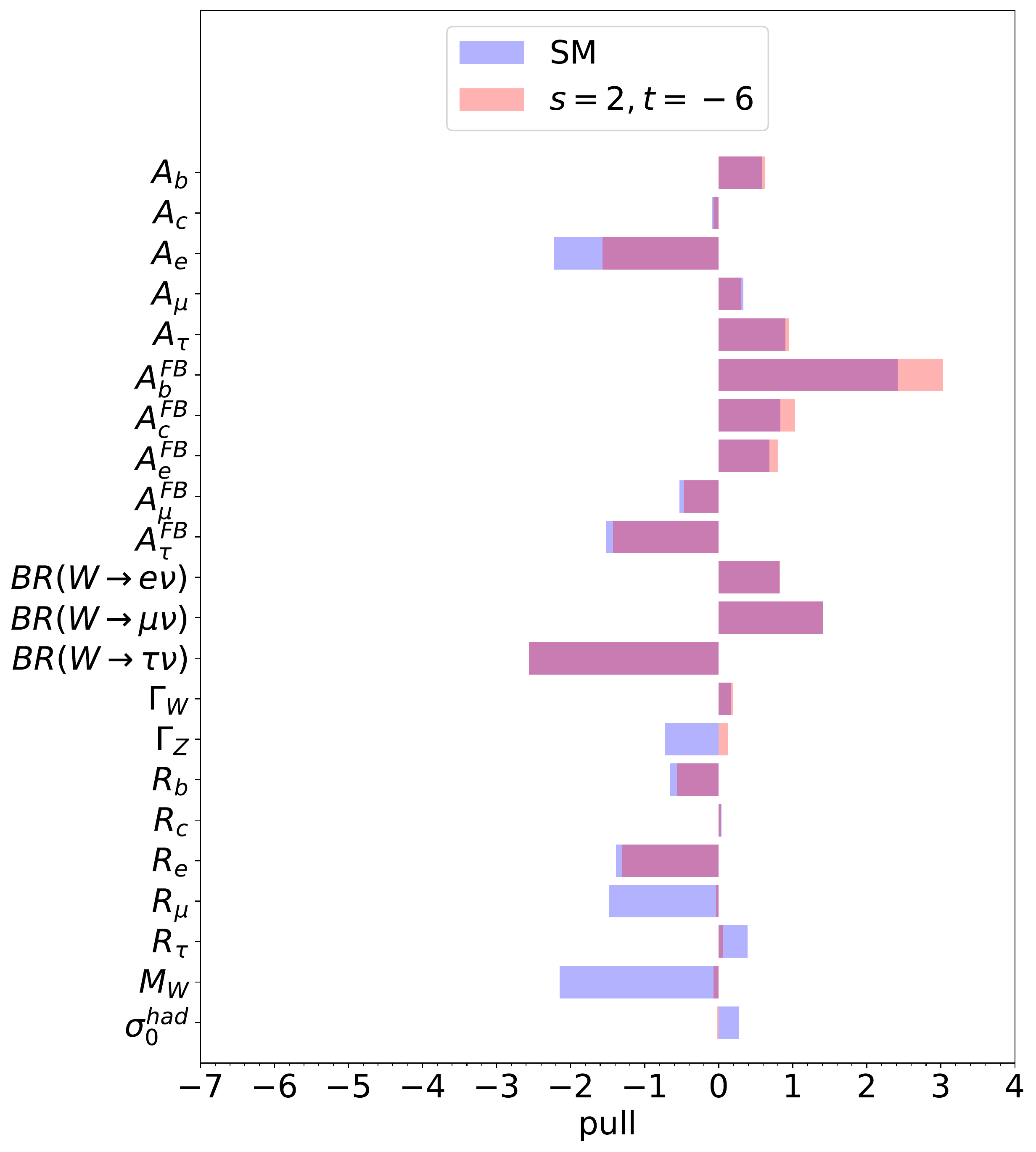}
  \includegraphics[width=0.5\textwidth]{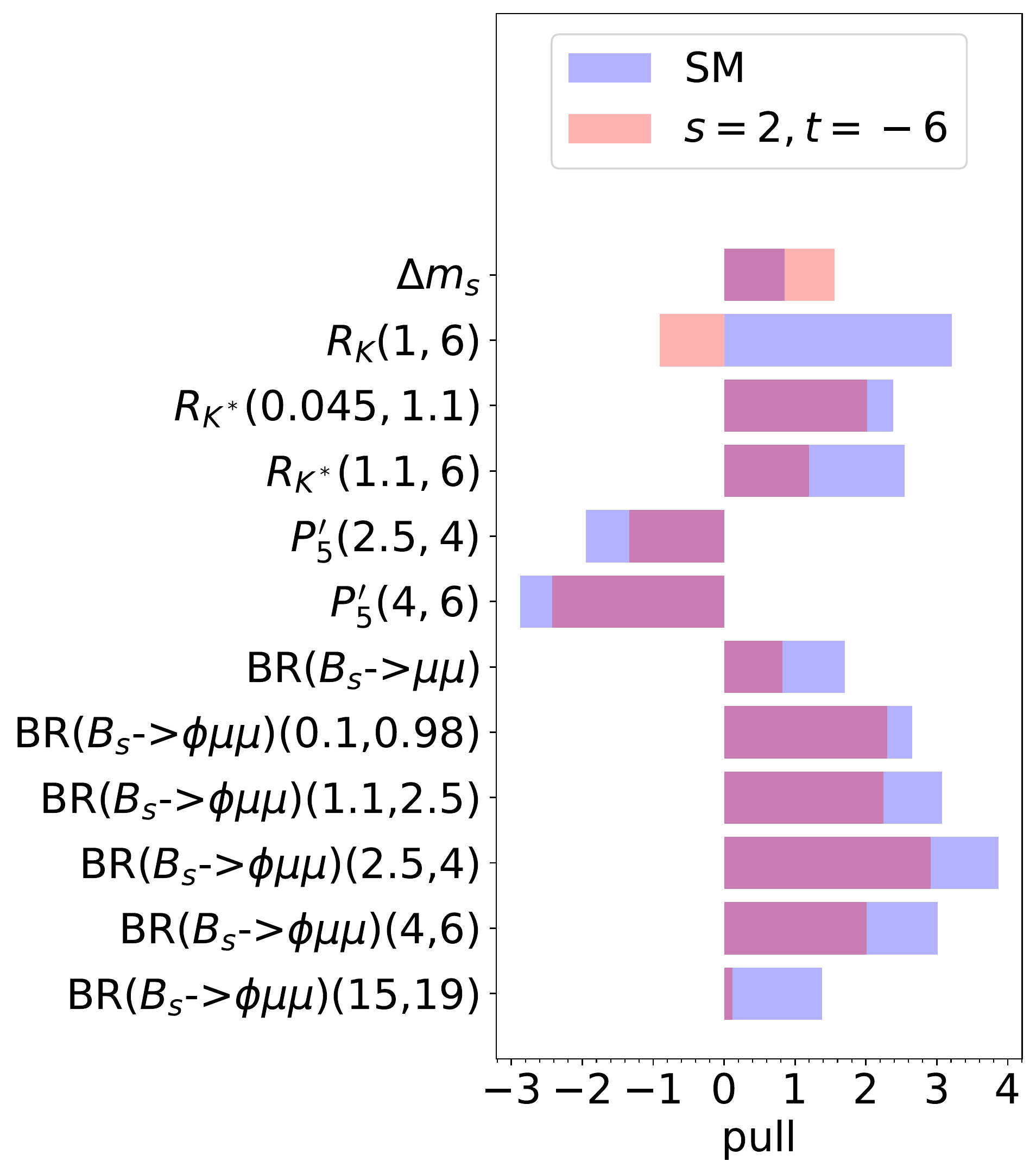}  
  \caption{\label{fig:pull_old}Pulls of interest for the default \smelli\ constraint
    on $M_W$ (i.e.\ excluding the recent CDF measurement) and $n=3$ as in (\ref{smelliMW}).
    In the
  left-hand panel, we display the EWPOs, whereas in the right-hand panel, we
  display selected observables of interest to $b \rightarrow s l^+l^-$
  anomalies.} 
  \end{figure}
We display the pulls for the combination of $M_W$ measurements that include the recent CDF II
determination in Fig.~\ref{fig:pull_new}. Several notable effects are evident:
for example, unsurprisingly $M_W$ itself is better fit, with a pull of $-1$. The observable $R_\mu$, the branching ratio
of the $Z^0$ boson to $\mu^+\mu^-$, has been increased by a fair amount; in the SM the
prediction was high by over 1$\sigma$, but in the $Z^\prime$
model with $t/s=-3$ it is less than one sigma too \emph{low}\footnote{Observables such as $R_\mu$ receive contributions from
$(C_{\phi l}^{(1)})^{22}$, $C_{\phi e}^{22}$. As a reference to
  Table~\ref{tab:WCs} and (\ref{impQ}-\ref{eq:charges}) shows, the sign of
  these contributions depends upon the sign of $t/s$. We have checked that
  the predictions for such observables do indeed go in the `wrong' direction
  for positive $t/s$, resulting in a less favourable fit.}.
We see that the forward-backward
asymmetry measured in $e^+e^-\rightarrow b \bar b$, $A_{FB}^b$, has a worse fit
in the $Z^\prime$ model. $A_e$, the left-right asymmetry in $e^+e^-\rightarrow
e^+e^-$, has a smaller pull in the $Z^\prime$ model. Overall, the quality of
fit to the EWPOs is fine: Table~\ref{tab:compp} shows that the $p$-value is
.29. The right-hand panel of Fig.~\ref{fig:pull_new} shows that many of the
selected observables of interest to $b-$meson decays are fit better than in
the SM save for the $B_s-\overline{B_s}$ mixing observable $\Delta m_s$,
which receives (fairly mild) positive corrections from the $Z^\prime$ contribution. 

We compare and contrast the fits \emph{including} the CDF II $M_W$ measurement
from Fig.~\ref{fig:pull_new}
with fits \emph{excluding} it in Fig.~\ref{fig:pull_old}. The most obvious
effect of excluding the CDF II $M_W$ measurement
is that the SM pull of $M_W$ of is only
2$\sigma$ when the CDF II $M_W$ measurement is excluded. Because the required
shift in $M_W$
is smaller, the relative effect upon the other 
EWPOs is smaller and the result is a good fit to EWPOs for $t/s=-3$: Table~\ref{tab:compp} reveals
the $p-$value in the fit to be .49. The fits in the $b-$observables on the
right-hand panel show a very similar pattern between Figs.~\ref{fig:pull_new}
and~\ref{fig:pull_old}. Essentially, $g_X$ is being fixed by the
EWPOs (and is being pulled by $M_W$ in particular), and then $\theta_{sb}$ is
fit to a value which fits the $b\rightarrow s \ell \ell$
anomalies. Here, many of the SM-discrepant observables relevant to the
$b-$anomalies receive a relative contribution from the $Z^\prime$~\cite{Allanach:2018lvl}
$
  \propto g_X^2 \sin 2 \theta_{sb}/M_X^2.
$
We see that the pull of $\Delta m_s$, which is a measure of the
  $B_s-{\overline B_s}$ mixing, decreases when one includes the CDF II $M_W$
  measurement. This is because the fit is pushed to a larger value of $g_X$ in
  order to fit the larger needed new physics contribution in $M_W$. In turn,
  in order to fit the $b-$anomalies, one requires a smaller value of
  $\theta_{sb}$ in order to keep the $\ZP$ contributions to them (\ref{eq:N})
  constant, in turn reducing the $\ZP$ contribution to $B_s-{\overline B_s}$ mixing.

The $p-$values and pulls for the $n=2$ fits are qualitatively similar to those
for $n=3$ and we neglect to present them here, noting that they are presented
in the ancillary information attached to the {\tt arXiv} preprint version of
this paper. Excluding all of the $M_W$ measurements except for the CDF II one,
the fits do not differ in significant details (the global $p-$values differ by
less than .02, for example) and we also relegate plots for them to the
ancillary information. 

\section{Conclusions \label{sec:conc}}
A $Z^\prime$ model where the SM is augmented by an additional, spontaneously
broken $U(1)$ gauge group can simultaneously fit both the CDF $M_W$ anomaly
and the $b\rightarrow s \ell \ell$ anomalies. The model retains the other
desirable properties of the $Y_3$ model on which it is
based~\cite{Allanach:2018lvl}: namely 
that it qualitatively explains a hierarchically heavy third generation of quarks
and small CKM angles. 
As
Fig.~\ref{fig:pvals} demonstrates, the log likelihood contribution from 
CDF's $M_W$ measurement is instrumental in picking out favoured values for the $U(1)_X$
quantum numbers of the fields. A particularly simple anomaly-free combination $Y_3
- 3 (B_3-L_3)$ has a high quality of fit (where the $\ell_3$ field is
subsequently identified with 
the left-handed \emph{muon} lepton doublet\footnote{Identifying $e_3$ with the right-handed muon
  field in addition also provides a similarly acceptable fit.}). We note that
for the combination $Y_3 - 3(B_3-L_3)$,
the `LFU' and `EWPO' classes of observable (both of which have small theoretical
uncertainties) each separately have a better quality of fit than the `quarks' class,
where there is more room for argument about the prediction and the size of the
theoretical uncertainty assigned. 

We have not developed details of the ultra-violet model, preferring instead to
begin by 
working with an effective theory with $SU(3)\times SU(2)_L \times U(1)_Y
\times U(1)_X$ gauge symmetry\footnote{The sub-Planckian Landau
  poles~\cite{Bause:2021prv} in $g_X$ may then be mitigated by new heavy (but
  sub-Planckian) states originating from a larger non-abelian symmetry that is
  spontaneously broken to SM$\times U(1)_X$.}. Light
family Yukawa couplings are expected to 
result from some non-renormalisable
operators, having integrated out heavy multi-TeV fermions that are vector-like
representations of the gauge group, for example. We
argue, following Ref.~\cite{aeon}, that it is premature to set the details of the model in the
ultra-violet more than we have, preferring instead to allow the data (such as
the new measurement of $M_W$) to inform the model building in a more
fundamental and vital way. Indeed, we have used it in the present paper to
select $U(1)_X$ charges under the symmetry group. 

In the coming years, the LHC experiments will provide valuable further
empirical measurements of $M_W$ and observables pertinent to the $b\rightarrow
s \ell \ell$ anomalies and Belle II data will also weigh in.
In the meantime, an obvious avenue of interest is from direct searches for the
$Z^\prime$. The most promising channels~\cite{Allanach:2021bbd}
are likely to be $pp \rightarrow
Z^\prime\rightarrow \mu^+ \mu^-$ with or without additional $b-$jets.
LHC and
LHC-HL sensitivity 
estimates are an obvious target for our models for future
study\footnote{Re-casts of LHC Run II $Z^\prime$ searches for similar models
  (e.g.\ the $Y_3$ and $B_3-L_3$ models)
indicate that the bounds are currently rather weak~\cite{Allanach:2021bbd}.}.

\section*{Acknowledgements}
This work has been partly supported by STFC HEP Theory Consolidated grant
ST/T000694/1. BCA thanks other members of the Cambridge Pheno Working Group
for useful discussions. 
JD is supported by the SNF under contract 200020-204428, and by the European Research Council (ERC) under the European Union’s Horizon 2020 research and innovation programme, grant agreement 833280 (FLAY).
We thank P Stangl for helpful communications about \smelli{} and A Crivellin
for useful comments on the manuscript. We further thank G Hiller, D Litim, T Steudtner, and especially T H\"ohne for enlightening discussions.
\bibliographystyle{JHEP}
\bibliography{Y3MW_draft}

\end{document}